\documentclass[12pt]{article}%
\usepackage{amsmath,latexsym}
\usepackage{graphicx}
\usepackage{amsmath}
\usepackage{amsfonts}
\usepackage{amssymb}%
\begin{document}

\title{{Noncommutative-geometry wormholes based on
   the  Casimir effect}}
   \author{
Peter K. F. Kuhfittig*\\  \footnote{kuhfitti@msoe.edu}
 \small Department of Mathematics, Milwaukee School of
Engineering,\\
\small Milwaukee, Wisconsin 53202-3109, USA}

\date{}
 \maketitle

\begin{abstract}\noindent
While wormholes are as good a prediction of
Einstein's theory as black holes, they are
subject to severe restrictions from quantum
field theory.  In particular, holding a
wormhole open requires a violation of the
null energy condition, calling for the
existence of exotic matter.  The Casimir
effect has shown that this physical requirement
can be met on a small scale, thereby solving
a key conceptual problem.  The Casimir effect
does not, however, guarantee that the
small-scale violation is sufficient for
supporting a macroscopic wormhole.  The
purpose of this paper is to connect the
Casimir effect to noncommutative geometry,
which also aims to accommodate small-scale
effects, the difference being that these
can now be viewed as intrinsic properties
of spacetime.  As a result, the noncommutative
effects can be implemented by modifying only
the energy momentum tensor in the Einstein
field equations, while leaving the Einstein
tensor unchanged.  The wormhole can therefore
be macroscopic in spite of the small Casimir
effect.     

\end{abstract}

\section{Introduction}\label{E:introduction}
Wormholes are handles or tunnels in spacetime
connecting widely separated regions of our
Universe or different universes in a multiverse.
In many ways, wormholes are as good a
prediction of Einstein's theory as black
holes, but they are subject to some severe
restrictions.  In particular, a wormhole can
only be held open by violating the null
energy condition, leading to what is called
``exotic matter" in Ref. \cite{MT88}.  This
outcome is not a conceptual problem in the
sense that exotic matter can be made in the
laboratory by means of the Caimir effect
\cite{hC48}.  Being a rather small effect,
it is not clear whether this is sufficient
for supporting a macroscopic traversable
wormhole.

Another area dealing with small effects is
noncommutative geometry, an offshoot of
string theory.  Here point-like particles
are replaced by smeared objects, an effect
that is often modeled using a Gaussian
distribution of minimal length $\sqrt{\beta}$.
Even though the existence of the small
parameter $\beta$ seems to echo the
Casimir effect, the noncommutative-geometry
background does not seem to prevent a
wormhole from being macroscopic.

The purpose of this paper is to examine
the possibility of combining these two
approaches.
%END OF SECTION

\section{Background}

\subsection{The Casimir effect}
To describe the Casimir effect, one normally
starts with two closely spaced parallel
metallic plates in a vacuum.  The plates can
be replaced by two concentric spherical
shells to preserve the spherical symmetry
preferred in a wormhole setting.  If $a$
is the separation between the shells, then
the pressure $p$ as a function of the
separation is \cite{rG19}
\begin{equation}\label{E:Casimir1}
   p(a)=-3\frac{\hbar c\pi^2}{720 a^4}
\end{equation}
and the density is
\begin{equation}\label{E:Casimir2}
   \rho_C(a)=-\frac{\hbar c\pi^2}{720a^4}.
\end{equation}
Here $\hbar$ is Planck's constant and $c$
is the speed of light.  Observe that the
equation of state has the form $p=\omega\rho$
with $\omega =3$.
%END OF SUBSECTION

\subsection{Noncommutatice geometry}
In recent years, string theory has become
ever more influential, as exemplified by
the realization that coordinates may
become noncommutative operators on a
$D$-brane \cite{eW96, SW99}.  The main
idea is that noncommutativity replaces
point-like particles by smeared objects
\cite{SS03, NSS06, NS10}, thereby eliminating
the divergences that normally appear in
general relativity.  As a result, spacetime
can be encoded in the commutator
$[\textbf{x}^{\mu},\textbf{x}^{\nu}]
=i\theta^{\mu\nu}$, where $\theta^{\mu\nu}$ is
an antisymmetric matrix that determines the
fundamental cell discretization of spacetime
in the same way that Planck's constant $\hbar$
discretizes phase space \cite{NSS06}.
According Refs. \cite{NSS06, mR11, RKRI, pK13},
the smearing can be modeled using a Gaussian
distribution of minimal length $\sqrt{\beta}$
instead of the Dirac delta function.  It is
shown in Refs. \cite{NM08, LL12} that an
equally effective way is to assume that the
energy density of a static and spherically
symmetric and particle-like gravitational
source is given by
\begin{equation}\label{E:rho1}
  \rho (r)=\frac{m\sqrt{\beta}}
     {\pi^2(r^2+\beta)^2}.
\end{equation}
The usual interpretation is that the
gravitational source causes the mass $m$ of
a particle to be diffused throughout the
region of linear dimension $\sqrt{\beta}$
due to the uncertainty.  In the next section,
we are going to be concerned with a smeared
spherical surface (referred to as the throat
of a wormhole).  So the smeared particle is
replaced by a smeared surface.  According
to Ref. \cite{pK20}, the energy density
$\rho_s$ is given by
\begin{equation}\label{E:rho2}
   \rho_s(r-r_0)=\frac{\mu\sqrt{\beta}}{\pi^2
   [(r-r_0)^2+\beta]^2},
\end{equation}
where $\mu$ now denotes the mass of the
surface.
%END OF SECTION

\section{Possible macroscopic wormholes}

Our first task in this section is to recall
the basic wormhole model proposed by Morris
and Thorne \cite{MT88},
\begin{equation}\label{E:line1}
ds^{2}=-e^{2\Phi(r)}dt^{2}+\frac{dr^2}{1-b(r)/r}
+r^{2}(d\theta^{2}+\text{sin}^{2}\theta\,
d\phi^{2}),
\end{equation}
using units in which $c=G=1$.  Here $b=b(r)$
is called the \emph{shape function} and
$\Phi=\Phi(r)$ is called the \emph{redshift
function}, which must be everywhere finite
to prevent the occurrence of an event horizon.
The spherical surface $r=r_0$ is called the
\emph{throat} of the wormhole, where $b(r_0)
=r_0$.  The shape function must also meet
the requirement $b'(r_0)<1$, called the
\emph{flare-out condition}, while $b(r)<r$
for $r>r_0$.  We also require that $b'(r_0)
>0$.

Returning now to Eq. (\ref{E:rho1}), the
energy density $\rho$ as a function of the
separation $r=a$ is
\begin{equation}
   \rho(a)=\frac{m\sqrt{\beta}}
        {\pi^2(a^2+\beta)^2}.
\end{equation}
According to Eq. (\ref{E:rho2}), in the
vicinity of the throat, i.e., whenever
$r-r_0=a$, we get
\begin{equation}
   \rho_s(a)=\frac{\mu\sqrt{\beta}}
        {\pi^2(a^2+\beta)^2}.
\end{equation}
So by Eq. (\ref{E:Casimir2}),
\begin{equation}\label{E:Casimir3}
   \frac{\mu\sqrt{\beta}}{\pi^2(a^2+\beta)^2}
   =|\rho_C(a)|=\frac{\hbar c\pi^2}
       {720a^4},
\end{equation}
thereby connecting the Casimir effect to
the noncommutative-geometry background.
Now, while the parameter $a$ may be small,
it is still macroscopic.  So we can assume
that $\beta=(\sqrt{\beta})^2\ll a^2$.
Being an additive constant, $\beta$ in the
denominator of Eq. (\ref{E:Casimir3})
becomes negligible.  So
\begin{equation}
   \sqrt{\beta}=\frac{\hbar c\pi^4}
       {720\mu}.
\end{equation}

Since $\hbar=1.0546\times 10^{-34}
\,\text{J}\cdot\, \text{s}$,
\begin{equation}\label{E:beta}
   \sqrt{\beta}=\frac{4.28\times 10^{-27}}
       {\mu}.
\end{equation}
Now recall that $\mu$ is the mass of the
throat $r=r_0$, a spherical surface of
negligible thickness.  This is an
idealization that is hard to quantify.
In practice, of course, we are dealing
with a thin shell that has a definite
mass $\mu$, thereby defining $\sqrt{\beta}$
in Eq. (\ref{E:beta}).  An alternative
approach is to give a direct physical
interpretation to the smearing effect by
letting $\sqrt{\beta}=a$.  Then Eq.
(\ref{E:Casimir3}) yields
\begin{equation}
   a=\frac{\hbar c\pi^4}{180\mu}
\end{equation}
and we can write
\begin{equation}
   a\mu =\frac{\hbar c\pi^4}{180},
\end{equation}
a fixed quantity.  So there are many
possible choices for $a$ and $\mu$.

Connecting the Casimir effect to
noncommutative geometry has some important\
consequences.  In particular, the
noncommutative effects can be implemented
in the Einstein field equations $G_{\mu\nu}
=\frac{8\pi G}{c^4}T_{\mu\nu}$ by modifying
only the energy momentum tensor, while
leaving the Einstein tensor intact.  The
reasons for this are discussed in Ref.
\cite{NSS06}: A metric field is a geometric
structure defined over an underlying
manifold whose strength is measured by
its curvature.  But the curvature is
nothing more than the response to the
presence of a mass-energy distribution.
Here it is emphasized in Ref. \cite{NSS06}
that noncommutativity is an intrinsic
property of spacetime rather than a
superimposed geometric structure.  So
it naturally affects the mass-energy
and momentum distributions, which, in
turn, determines the spacetime curvature,
thereby explaining why the Einstein
tensor can be left unchanged.  As a
consequence, the length scales can be
macroscopic.

A final issue to be addressed is the
large radial tension at the throat of
a Morris-Thorne wormhole.  First we
need to recall that the radial tension
$\tau(r)$ is the negative of the radial
pressure.  According to Ref. \cite{MT88},
the Einstein field equations can be
rearranged to yield
\begin{equation}
   \tau(r)=\frac{b(r)/r-2[r-b(r)]\Phi'(r)}
   {8\pi Gc^{-4}r^2}.
\end{equation}
It follows that the radial tension at the
throat is
\begin{equation}\label{E:tau}
  \tau(r_0)=\frac{1}{8\pi Gc^{-4}r_0^2}\approx
   5\times 10^{41}\frac{\text{dyn}}{\text{cm}^2}
   \left(\frac{10\,\text{m}}{r_0}\right)^2.
\end{equation}
In particular, for $r_0=3$ km, $\tau$ has
the same magnitude as the pressure at the
center of a massive neutron star.  This is
rather hard to explain if we are not
actually dealing with neutron matter.  It
is shown in Ref. \cite{pK20}, however, that
a noncommutative-geometry background can
indeed account for the large radial tension.
%END OF SECTION

\section{Conclusion}

While the possible existence of wormholes
is a consequence of Einstein's theory, such
wormholes would be subject to severe
restrictions from quantum field theory.  In
particular, holding a macroscopic wormhole
open requires a violation of the null energy
condition, calling for the existence of
exotic matter in the vicinity of the throat.
The Casimir effect has shown that such a
violation can be produced in the laboratory,
thereby eliminating a major conceptual problem.
So the real question is whether enough exotic
matter could be produced  to sustain a
macroscopic wormhole.  The purpose of this
paper is  to connect the Casimir effect to
noncommutative geometry, which also involves
small effects thanks to the parameter $\beta$
in Eq. (\ref{E:rho2}).  Here the gravitational
source causes the mass $\mu$ of a surface to
become diffused due to the uncertainty.  The
result is a wormhole whose throat is a smeared
surface.  The subsequent connection to the
Casimir effect yields a physical interpretation
of the length $\sqrt{\beta}$, which is a
measure of the uncertainty.  Apart from this,
the noncommutative effects can be implemented
in the Einstein field equations by modifying
only the energy momentum tensor, while
leaving the Einstein tensor unchanged.  We
conclude that the wormhole can indeed be
macroscopic in spite of the small Casimir
effect.  Finally, the noncommutative-geometry
background can account for the enormous
radial tension that is characteristic of
moderately-sized Morris-Thorne wormholes.

\end{document}